\documentclass[a4]{article}

\usepackage{graphicx}
\usepackage{hyperref}
\usepackage{amsmath}
\usepackage{cite}
\usepackage{cuted}
\usepackage{capt-of}
\usepackage{color}
\usepackage{bm}
\usepackage[normalem]{ulem}

\newcommand{\bfm}[1]{\textbf{#1}}

\newcommand{\dif}{{\rm d}}
\newcommand{\dvol}{{\rm d}^3\bfm{r}}

\newcommand{\dsec}{{\rm d}^2\bfm{r}_2}

\newcommand{\vJ}{\bfm{J}}
\newcommand{\vM}{\bfm{M}}
\newcommand{\vE}{\bfm{E}}
\newcommand{\vB}{\bfm{B}}
\newcommand{\vA}{\bfm{A}}
\newcommand{\vH}{\bfm{H}}

\newcommand{\ve}{\bfm{e}}
\newcommand{\vG}{\bfm{G}}
\newcommand{\vr}{\bfm{r}}

\newcommand{\vDJ}{\Delta\bfm{J}}
\newcommand{\vDA}{\Delta\bfm{A}}

\newcommand{\figwidth}{8.8 cm}

\begin{document}

\title{AC loss modelling in a 2 MW-class REBCO high temperature superconducting motor for hydrogen-electric aircraft} 

\author{Enric Pardo$^{1,*}$, Alexandre Colle$^{2}$, R{\'e}mi Dorget$^{2}$,\\
Mohamed Essam Ahmed$^{2}$ \\
$^1$Institute of Electrical Engineering, Slovak Academy of Sciences,\\
Bratislava, Slovakia\\
$^2$AIRBUS UpNext, Toulouse, France.\\
$^*$ Author to whom correspondence should be\\
addressed (enric.pardo@savba.sk).
}

\markboth{IEEE Transactions on Transportation Electrification}%
{E Pardo \MakeLowercase{\textit{et al.}}: AC loss modelling in a 2 MW-class REBCO motor ...}

\maketitle

\begin{abstract}
High temperature superconducting motors are very promising for hydrogen-electric aircraft thanks to their high specific power, specific torque, and efficiency. High temperature superconductor REBCO offer high cryogenic flexibility, but a stator made of REBCO tapes could present high AC loss. Although stacking effect reduce AC loss, it could be compromised by imperfections, such as winding misalignment and tape inhomogeneity. Therefore, it is needed to know whether the AC loss is acceptable in realistic REBCO stators. This article analyses the AC loss in a REBCO propulsion motor for aviation that takes these imperfections into account. For this purpose, we developed our own fast and accurate numerical model, which considers the highly nonlinear screening currents in the superconductor into account. This work studies a motor of around 2 MW power with REBCO stator coils with 27 parallel tapes as conductor and a permanent-magnet rotor. We consider several electric coupling scenarios of the multi-tape conductor. We also analyze the effect of finite tape-to-tape resistances at the terminals. We have found that the AC loss for the whole motor in the most realistic coupling scenario represents less than 0.018 \% of the rated power. Misalignments and tape degradation at the edges of up to 100 $\mu$m only increase AC loss by up to around 20 \%. Therefore, motors with REBCO  superconductors in the stator are feasible for aircraft propulsion.

{\bf Keywords:} Hydrogen electric aircraft, superconducting motors, REBCO high temperature superconductors, AC losses, computer modeling. 
\end{abstract}

\section{Introduction}

\begin{figure}[tbp]
{\includegraphics[trim=0 -4 0 0,clip,width=\figwidth]{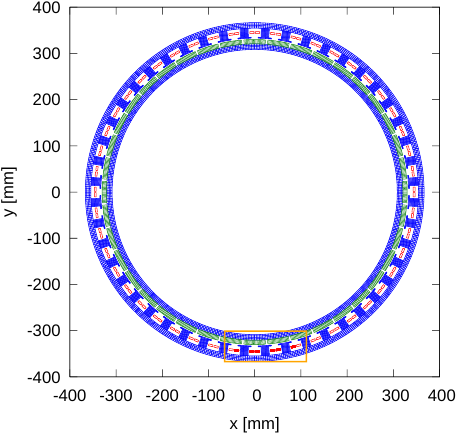}}\\%
{\includegraphics[trim=0 -4 0 0,clip,width=\figwidth]{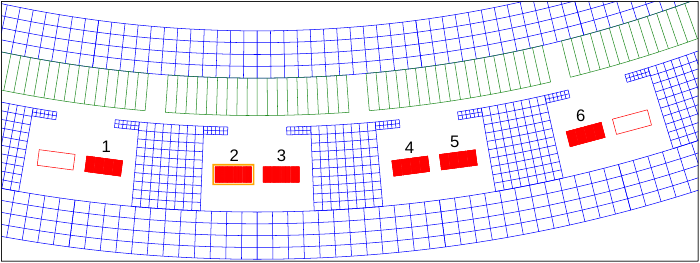}}\\%
{\includegraphics[trim=0 -4 0 0,clip,width=\figwidth]{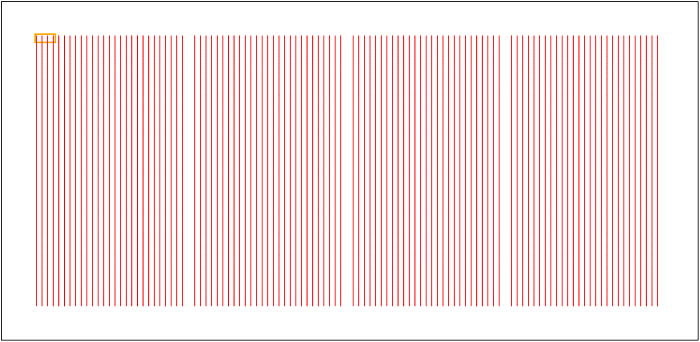}}\\%
{\includegraphics[trim=0 0 0 0,clip,width=\figwidth]{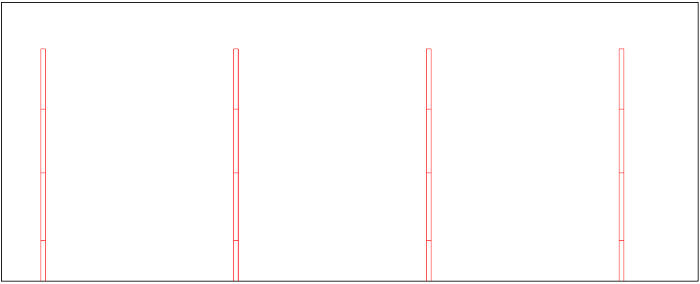}}%
\caption{Superconducting motor configuration in the present study with zooms in several parts (orange frame delimits the area of the next zoom). Modeling is from the scale of 1 m for the whole motor down to around 1 $\mu$m for the tape thickness. Wire frames in red, blue, and green are for the superconducting layers, iron yoke and permanent magnets, respectively. \label{f.motor}}%
\end{figure}

Commercial aviation is a growing source of greenhouse effect emissions, such as CO$_2$, NO$_x$ and other emissions that cause undesirable contrails \cite{hydrogen_aviation}. Hydrogen-electric aircraft are very promising, since they do not emit CO$_2$, if hydrogen is produced by renewable sources. On board electricity could be generated either by a generator with a hydrogen combustion turbine or by hydrogen fuel cells. The latter is the cleanest option, since it generates neither NO$_x$ nor contrails. In addition to low emissions, hydrogen-electric aircraft can benefit from distributed electric propulsion, which highly reduces drag, and hence energy consumption \cite{gohardaniAS2011PAS, barzkarA2020IEA}. This is because it is possible to place a relatively high number of electric propulsion motors on aircraft wings or body without significant loss of motor efficiency, while that is not the case for classical combustion turbines. Indeed, combustion turbofans need to be large in order to be efficient (several 10's of MW), but electric motors stay highly efficient when downscaling to a few MW per motor.

Superconducting electric motors for aircraft propulsion present several advantages compared to their normal conducting counterparts, such as higher specific power and higher efficiency \cite{haranKS2017SST, sayedE2021TTE, wangY2023Ene}. It is expected that only superconducting motors can reach the demanding requirement of 25 kW/kg for certain types of airliners \cite{FlyZero2022roadmap}. Actually, superconducting motors are only one component of the electric powertrain. A fully cryogenic powertrain, which also contains superconducting power-transfer cables and a cryogenic motor-control unit is expected to further reduce weight compared to conventional electric powertrains. Powertrains with superconductors are especially promising for hydrogen-electric aircraft, since the liquid hydrogen tank represents a practically unlimited heat sink. Although liquid hydrogen will be the ultimate heat sink, a secondary circuit of helium gas is the most suitable option to cool the cryogenic components, for safety \cite{ybanezL2022JCS, cryoprop2025, caughleyA2024Aer}.

At present, there is a high interest in the topic of superconducting motors for aircraft electric propulsion \cite{haranKS2017SST, manolopoulosCD2018IES, grilliF2020JCS, wengF2020SSTa, kalsiSS2023IES, gaoY2023PhC, miyazakiH2024IES, dorgetR2025IES, mellerudR2025IES, balachandranT2026IES}. The highest specific power could be achieved by a fully superconducting motor, and hence both stator and rotor contain superconducting coils. Intermediate technological steps are motors with either superconducting rotor or stator. Motors with superconductor only in the stator present simpler cryogenics than those with superconducting rotor, since no rotating vacuum components are required. However, superconducting stators experience significant power heat under alternating currents (AC loss). This AC loss needs to be kept as low as possible, since heat extraction at cryogenic temperatures could be complex. Insufficient heat extraction may cause electrothermal quench, which may damage the superconductor. One strategy to reduce the AC loss is to divide the superconducting wire or tape into filaments and then transpose by appropriate twisting or cabling. This has been the approach for stators based on multi-filamentary MgB$_2$, for instance \cite{manolopoulosCD2018IES, kalsiSS2023IES}. Another way that was proposed in \cite{pardoE2019IES, grilliF2020JCS, pardoE2019EUCAS} is to use the stacking effect in superconducting tapes \cite{pardoE2003PRB}, such as REBCO ($RE$Ba$_2$Cu$_3$O$_{7-x}$ where $RE$ is a rare earth, typically Y, Gd or Eu). REBCO high temperature superconductors are very promising for cryogenic powertrains because of their high cryogenic flexibility, since the material retains good electric properties (such as critical current) up to around 50 K, while liquid hydrogen boils at 20 K. On the other side, powertrains based on MgB$_2$ present much more challenging cryogenics, since their critical current falls when increasing the temperature a few degrees above 20 K. 

AC loss prediction in REBCO superconducting motors requires computer modeling, since analytical estimations for single tapes neglect stacking effects, and hence they incur to erroneous predictions by orders of magnitude. AC loss modeling of superconducting motors typically use Finite Element Methods (FEM) for the $T-A$ formulation \cite{benkelT2020IES, huberF2022SST}. However, previous works used a combination of the Minimum Electromagnetic Entropy Production (MEMEP) \cite{pardoE2015SST} with FEM in $A$ formulation \cite{pardoE2019IES}, which enabled fast and accurate parametric studies.

In this work, we present a thorough study of the AC loss in a motor with a REBCO stator (figure \ref{f.motor}), showing that the generated heat is tolerable. In particular, we analyze the impact on AC loss of issues that could compromise the stacking effect, such as tape misalignment and degradation of critical current at the edges. We also study several types of electric connection between tapes (electric coupling). For this purpose, we developed a novel numerical model that combines the Minimum Electro Magnetic Entropy Production (MEMEP) for the superconductor interacting with the nonlinear magnetic iron yoke and the rotor, and hence no combination with FEM is necessary. This method obtains details of the current density, which presents strong screening currents, and accurately computes the AC loss.

\section{Configuration}

\subsection{Motor with REBCO superconducting stator}

\begin{table}[tbp]
\caption{Superconducting motor parameters}\label{t.param}
\center
\begin{tabular}{ll}
\hline 
\hline 
Motor power & $\sim$ 2 MW \\
Rotation speed & 1000 rpm \\
Operating temperature & 40 K \\
Active length & 335 mm \\
Superconductor material & REBCO tape$^*$ \\
Number of coils & 48 \\
Number of turns & 4 \\
Number of tapes in parallel & 27 \\
Tape width & 4 mm \\
Superconducting layer thickness & 1 $\mu$m \\
Stator current amplitude & 1308.15 A \\
Stator current frequency & 333 Hz \\
Maximum stator voltage amplitude & $\sim$ 400 V \\
Iron yoke material & Cobalt-steel alloy \\
Magnet magnetization ($\mu_0M_{PM}$) & 1.2 T \\
Number of magnets & 40 \\
\hline
\hline
\end{tabular}

\vspace{2mm}
$^*$ We assume the $J_c(B,\theta,T)$ dependence of SuperOx YBCO 2G HTS 2021 from \cite{robinson_data}. Present REBCO usually experiences higher $J_c$.  
\end{table}

In this article, we consider the motor configuration of figure \ref{f.motor} with parameters in table \ref{t.param}. This is a radial-flux motor with a REBCO superconducting stator and a permanent-magnet rotor. Each stator coil is made of only 4 turns in order to reduce the inductance, which allows operation of relatively low voltage amplitudes (around 400 V). The phases of the current in the stator coils are the following: -A,A,B,-B,-C,C,A,-A,-B,B,C,-C and so on in the anticlockwise direction.

The conductor consists on many REBCO tapes in parallel, 27, in order to enable high current capacity. Indeed, the operation current, 1308.15 A amplitude (925 A rms), is well below the critical current. The purpose is twofold. First, this reduces the AC loss by stacking effect \cite{pardoE2003PRB, pardoE2019IES, grilliF2020JCS, pardoE2020ASC}, and hence shielding both the magnetic field from the rotor and self-field from the stator. Second, this minimises the risk of electrothermal quench, which occurs when the current in any tape is well above its critical current. All turns are electrically insulated, in order to avoid high loss due to transverse current in no-insulation or metal-insulation coils \cite{pardoE2024SSTa}. The baseline design is that all tapes are insulated along their length, but we also regard the configuration of soldered tapes, for completeness (section \ref{s.coupl}).

In this work, we focus on the AC loss generated at the superconducting layer, which we assume of 1 $\mu$m thickness (figure \ref{f.motor}). Then, we neglect the AC loss contribution from the metal layers in the superconducting tape (Hastelloy substrate, silver, and copper layers).

\subsection{Electric coupling between filaments}

For multi-tape cables there could appear coupling currents that flow between superconducting tapes through normal metals that join them. These coupling currents can highly increase the AC loss because of both Joule effect in the joining metals and increased AC loss in the superconductor.

\subsubsection{Simplified coupling configurations}
\label{s.coupl}

\begin{figure}[tbp]
{\includegraphics[trim=0 0 0 0,clip,width=\figwidth]{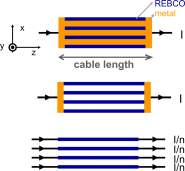}}%
\caption{Simplified coupling configurations (``coupled'', ``coupled-at-ends'', and ``uncoupled'' from top to bottom), where $y$ and $z$ are the directions of the tape width and length, respectively. Above, $I$ and $n$ are the coil current and number of parallel tapes, respectively. We assume that the metal presents zero resistivity. \label{f.coupl}}%
\end{figure}

In a first stage, we consider the 3 simple electric coupling configurations in figure \ref{f.coupl}. The ``coupled'' case (figure \ref{f.coupl}(top)) assumes that all tapes are soldered to each other along their length by a metal with zero resistivity. The ``coupled-at-ends'' configuration (figure \ref{f.coupl}(mid)) considers that all tapes are insulated from each other along the length but they are connected at the ends by a metal with zero resistivity. This is a realistic configuration, since tapes need to be connected to each other at the current leads. Finally, we assume the idealised ``uncoupled'' configuration (figure \ref{f.coupl}(bottom)), where all tapes are insulated from each other and we impose that all tapes transport the same net current, and hence there are no coupling currents.

One can model these configurations as follows. For coupled, the current distributes freely among all tapes and this distribution can be different in each turn. For coupled-at-ends, the current also distributes freely but, since all tapes are insulated along their length, the coil current, $I$, distributes in the same proportion for all turns. The uncoupled case just assumes that current $I$ distributes equally among all tapes. All these configurations assume no AC loss in the metals joining the tapes because their resistivity vanishes, and hence any AC loss increase compared to the uncoupled case will be due to higher AC loss in the superconductor.

\subsubsection{Coupling by finite resistances at terminals}

\begin{figure}[tbp]
\center
{\includegraphics[trim=0 0 0 0,clip,height=6 cm]{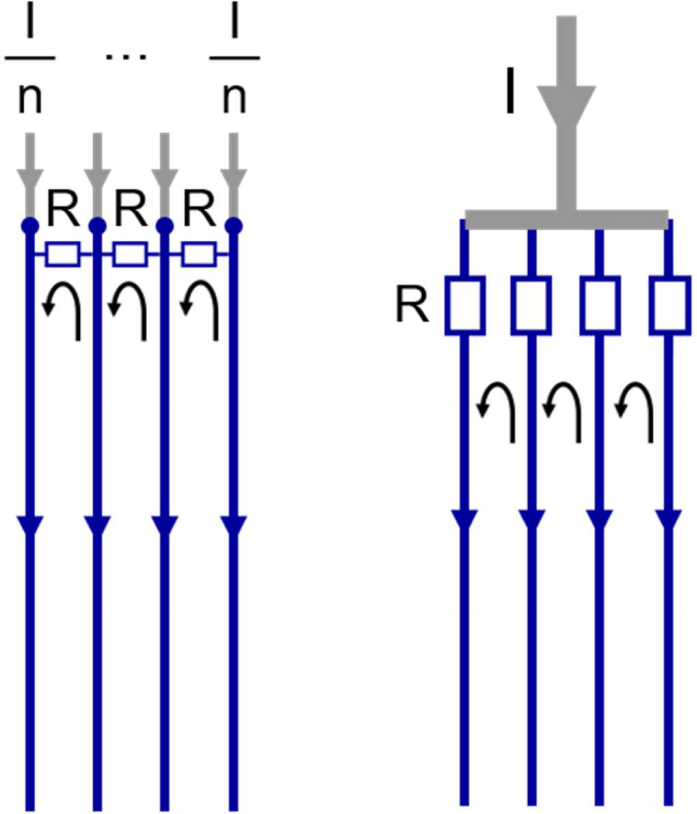}}%
\caption{Coupling configurations considering finite resistances between the parallel tapes at the coil ends. Left: intermediate configuration between ``ucoupled'' and ``coupled-at-ends'' of figure \ref{f.coupl}. Right: Realistic coupling at the terminals. \label{f.couplR}}%
\end{figure}

In this work we also analyse the effect of finite resistances at the terminals. These directly generate AC loss and they cause a superconductor loss increase that is in between the uncoupled and coupled-at-ends case. Here, we assume that the ``tape soldering'' configuration of figure \ref{f.couplR}(left). Although the ``current lead'' case of figure \ref{f.couplR}(right) is more realistic, we chose to study the ``tape soldering'' connection. The reason is to analyse in detail the transition from ``uncoupled'' to ``coupled-at-ends'' with decreasing the solder resistance, $R_s$. Then, the computed AC loss corresponds to the AC contribution, being the DC contribution in the ``current lead'' case straightforward. The AC loss contribution (total power loss minus the DC loss) for the two types of connection of figure \ref{f.couplR} will be the closest when the resistance of the current flowing from opposite tapes will be the same. This is for 
\begin{equation} \label{Rl}
R_l=nR_s/2,
\end{equation} 
where $R_l$ is the resistance at the current leads and $n$ are the number of tapes. When there are only significant coupling currents between the two opposite tapes, both connections result in the same AC loss.

\section{Numerical modelling method}

In order to reduce computing time while keeping high accuracy, we consider only the details of 3 superconducting coils (6 half-coils), as shown in figures \ref{f.motor}(a,b). The solid red rectangles there are superconducting half-coils, while the open red frames are coils where current density is assumed uniform. In this way, we have two full slots with superconducting coils. We also consider one half-coil that is beyond these two full slots in each direction because we need to consider full coils in order to take the coupled-at-ends configuration into account. Since for this magnet the time-average loss per cycle at each slot becomes periodic in space by pairs of coils, the total time-average loss is that of the two slots with two superconducting half-coils times half the number of slots. This assumption is accurate because the interaction of screening currents between two half-coils is only relevant for those within the same slot. Indeed, we have made computations with up to 10 half-coils and we have seen that the improvement in loss accuracy is negligible compared to 6 half-coils (two full slots).

Modelling is from the scale of 1 m for the whole motor down to around 1 $\mu$m for the tape thickness (figure \ref{f.motor}), and hence the model is multi-scale.

\subsection{Superconductor and normal conductor modelling}
\label{s.scmod}

\begin{figure}[tbp]
\center
{\includegraphics[trim=3 1 10 5,clip,width=\figwidth]{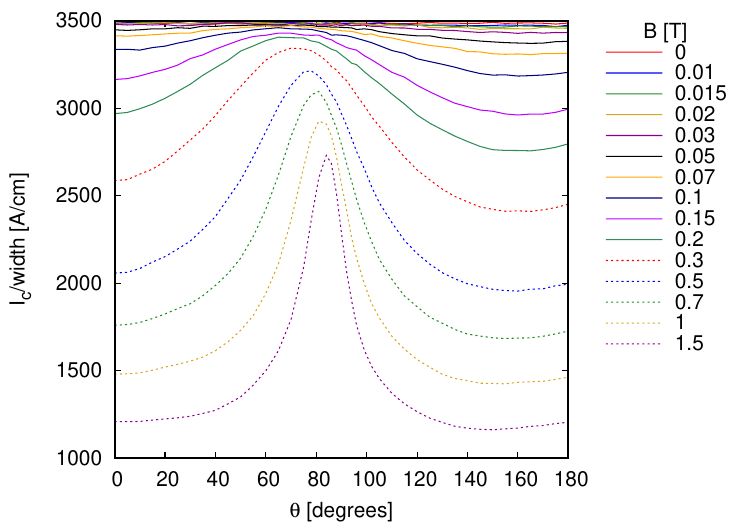}}%
\caption{Dependence of the tape critical current, $I_c$, as a funcion of the applied magnetic field, $B$, and its angle with the tape normal at 40 K. Data for SuperOx YBCO 2021 from \cite{robinson_data} (only curves for up to 1.5 T are shown). The critical current density is $J_c=I_c/(wd)$, where $w$ and $d$ are the superconductor width and thickenss, respectively \label{f.JcBth}}%
\end{figure}

We model the superconductor and conducting materials by means of the Minimum Electromagnetic Entropy Production (MEMEP) method \cite{pardoE2015SST, pardoE2017JCP}. This method enables to model any material with any physical nonlinear relation between the current density, $\vJ$, and the electric field, $\vE$. For the superconductor, we assume that the $\vE(\vJ)$ relation follows a power law
\begin{equation} \label{EJ}
\vE(\vJ)=E_c \left ( \frac{|\vJ|}{J_c} \right )^{n_c} \frac{\vJ}{|\vJ|},
\end{equation}
where $E_c=10^{-4}$ V/m, $J_c$ is the critical current density and $n_c$ is the power law exponent. In general, both $J_c$ and $n_c$ depend on the magnitude of the magnetic flux density, $B$, its angle with the tape surface, $\theta$, and the temperature, $T$. We use the input $J_c(B,\theta,T)$ data from the Robinson Research Insititute database \cite{robinson_data} for the SuperOx YBCO 2021 tape (see figure  \ref{f.JcBth}). Although we may also use $n_c(B,\theta,T)$ provided in that database, in this work we assume $n_c=30$ for simplicity. Any normal conducting parts of the tape follow Ohm's law
\begin{equation} \label{Ohm}
\vE(\vJ)=\rho_n\vJ ,
\end{equation}
where $\rho_n$ is the normal resistivity of the metal. This has been considered to model the current in the resistances at the coils ends (figure \ref{f.couplR}).

In brief, MEMEP solves $\vJ(\vr,t)$ for any constitutive $\vE(\vJ)$ relation by solving the following equation
\begin{equation} \label{EAphi}
\vE(\vJ)=-\frac{\partial \vA[\vJ]}{\partial t}-\frac{\partial \vA_a}{\partial t}-\nabla\phi,
\end{equation}
where the vector potential in Coulomb's gauge ($\nabla\cdot\vA=0$ and $|\vA|\to 0$ at $|\vr|\to\infty$) has been decomposed into the contribution of external sources that are independent on $\vJ$, $\vA_a$, and the contribution of $\vJ$ \cite{pardoE2023book}:
\begin{equation}
\vA[\vJ](\vr)=\frac{\mu_0}{4\pi}\int_\Omega\dvol' \frac{\vJ(\vr')}{|\vr-\vr'|},
\end{equation}
where $\mu_0$ is the void permeability and $\Omega$ is the region where there is current density. In (\ref{EAphi}), $\phi$ is the electrostatic potential. Appendix \ref{s.MEMEP} details how MEMEP solves (\ref{EAphi}) in order to compute the time evolution of $\vJ(\vr,t)$.

Once $\vJ(\vr,t)$ is known, we compute its contribution to $\vB(\vr,t)$ in the superconductor and iron yoke based on pre-calculated interaction matrices using the Biot-Savart law. This also enables to calculate $\vB$ at regions in the air, such as the stator-rotor gap, without a significant increase in computing time.

The advantages of MEMEP compared to conventional Finite Element Methods (FEM) is that MEMEP solves $\vJ(\vr,t)$ at the superconductor (or normal conductor) only, avoiding to spend degrees of freedom in the air. For infinitely long problems, degrees of freedom are further minimized because $\vJ$ has only component int the $z$ direction. In addition, MEMEP is highly parallelized, which enables to exploit multi-core CPUs. Thanks to this, MEMEP is more time efficient than commercial codes \cite{ainslieM2020SST}. Finally, MEMEP simplifies parametric studies that change the superconductor geometry, since we can use the same mesh for the superconductor. 

Compared to \cite{pardoE2015SST, pardoE2017JCP}, we programmed new software for 2D cross-sectional problems, where we assume that the conductor (or coil) length is much larger than its width. This enables to compute the current density, $\vJ$, and power AC loss per unit volume, $p$.	

\subsection{Iron yoke modelling}

\begin{figure}[tbp]
{\includegraphics[trim=0 0 0 0,clip,width=\figwidth]{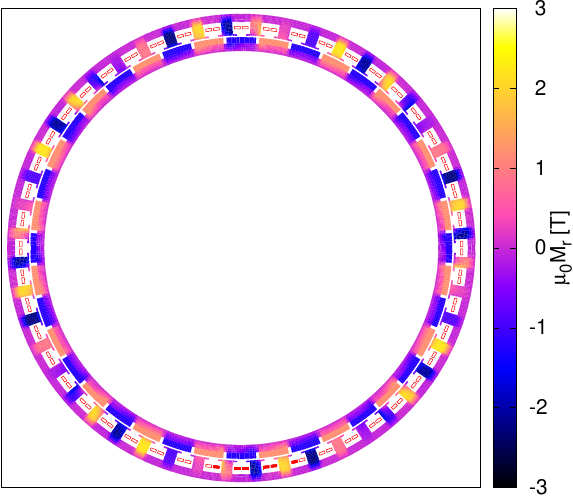}}\\%
{\includegraphics[trim=0 0 0 0,clip,width=\figwidth]{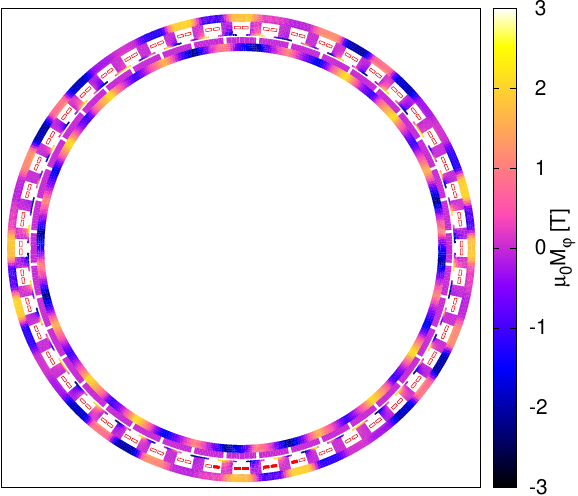}}%
\caption{Example of the solution of the magnetization, $\vM$, in the iron yoke and the permanent magnets for the radial component (top) and the angular component (bottom). \label{f.M}}%
\end{figure}

\begin{figure}[tbp]
{\includegraphics[trim=0 0 0 0,clip,width=\figwidth]{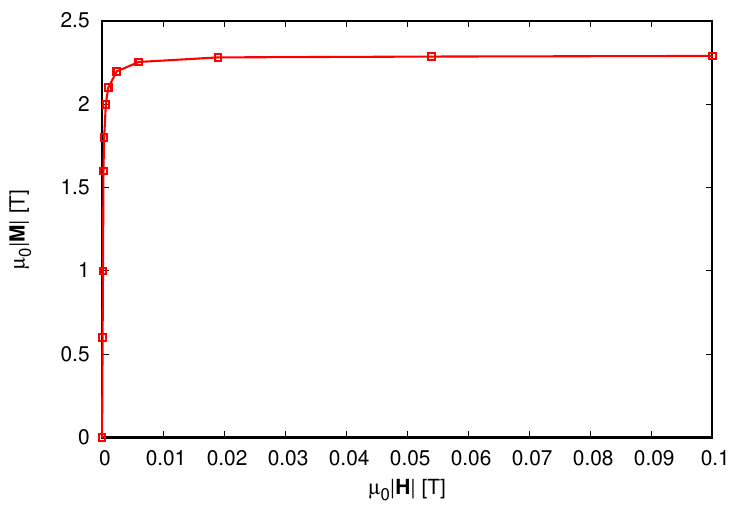}}\\%
{\includegraphics[trim=0 0 0 0,clip,width=\figwidth]{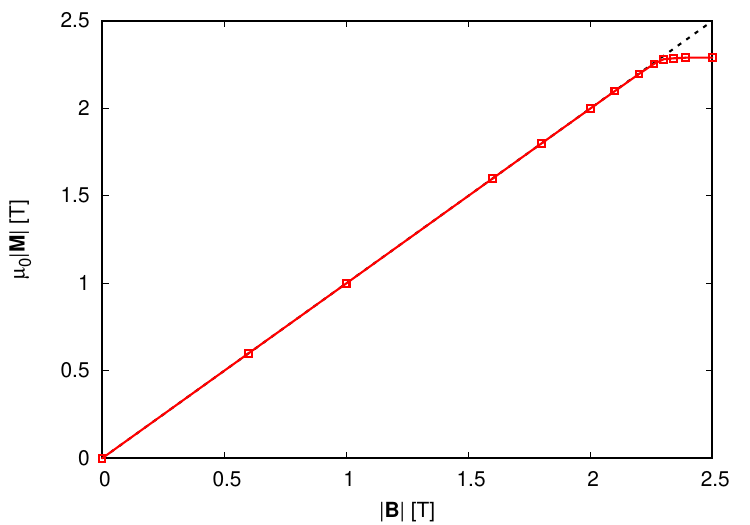}}%
\caption{Example of $|\vM|-|\vH|$ and $|\vM|-|\vB|$ relations, using data from Cobalt-steel alloy given by the provider. Here, we have assumed that the material is isotropic. The dash line in the $|\vM|-|\vB|$ relation is a line with slope 1 as a guide for the eye. \label{f.MB}}%
\end{figure}

The magnetization, $\vM$, in the iron, which is assumed as reversible, is solved as follows.

First, we obtain a relation between $\vM$ and the magnetic flux density, $\vB$, as $\vM=\bfm{G}(\vB)$ from the measured relation between $\vB$ and the magnetic field, $\vH$. The measured $\vB-\vH$ relation can be written as $\vH$ being a function of $\vB$, $\vH(\vB)$. From the definition of $\vH$, $\vB=\mu_0(\vH+\vM)$, we obtain
\begin{equation} \label{MB}
\vM=\frac{1}{\mu_0}(\vB-\mu_0\vH(\vB)) \equiv \bfm{G}(\vB).
\end{equation}
For isotropic materials, $\vM$, $\vB$, and $\vH$ are parallel to each other. Then,
\begin{equation}
|\vM|=\frac{1}{\mu_0}\left( 1-\frac{1}{\mu_r(|\vB|)} \right )|\vB|,
\end{equation}
where $\mu_r=|\vB|/(\mu_0|\vH|)$ is the relative permeability. Alternatively, we may use a $\vM(\vH)$ relation as $\vM=(\vB(\vH)-\mu_0\vH)/\mu_0$ instead of the $\vM(\vB)$ relation of (\ref{MB}). However, a model using $\vM(\vB)$ is more robust than using $\vM(\vH)$ because a small error in $\vB$ has only a small impact in $\vM$, while a small error in $\vH$ can have a high impact in $\vM$ when $|\vH|$ is small (see figure \ref{f.MB}).

Next, we use $\vB$ that has two contributions, one from $\vM$ and another from external sources, $\vB_a$, and hence
\begin{equation}
\vB=\vB[\vM]+\vB_a,
\end{equation}
where $\vB[\vM]$ for general three dimensional (3D) shapes is
\begin{equation}
\vB[\vM](\vr)=\frac{\mu_0}{4\pi}\int_\Omega\dvol'
\frac{\left ( \nabla'\times\vM(\vr') \right )\times (\vr-\vr')}{|\vr-\vr'|^3},
\end{equation}
where $\Omega$ is the region where there is magnetization. For infinitely long shapes under transverse applied field, as is the case of this article, $\vB[\vM]$ follows
\begin{equation}
\vB[\vM](\vr_2)=\frac{\mu_0}{2\pi}\int_\Omega\dsec' 
\frac{\left ( \nabla'\times\vM(\vr'_2) \right )\times (\vr_2-\vr'_2) }{|\vr_2-\vr'_2|^2},
\end{equation}
where $\vr_2$ is the position vector in the object cross-section, which only has two components. In order to evaluate $\vB[\vM]$, we discretize $\vM(\vr)$ in space assuming uniform $\vM$ in each element (blue trapezoids in figure \ref{f.motor}). Then, we compute the interaction matrix between each component of $\vM$ at any element $i$ with each component of $\vB[\vM]$ at any element $j$. The interaction matrix enables a fast computation of $\vB[\vM]$ at any element for any update of $\vM$ at any other element.

Finally, we solve $\vM$ at all elements iteratively. We start with $\vM=0$ (and consequently $\vB[\vM]=0$) everywhere. Afterwards, $\vM$ at iteration $k+1$ is obtained from $\vM$ at the previous iteration, $k$, as
\begin{equation}
\vM_{k+1}=\vG(\vB[\vM_{k}]+\vB_a).
\end{equation}
We keep iterating until the difference in any component of $\vM$ is below a certain tolerance. This method is simple, fast, accurate, and fully parallelizable. 

\subsection{Interaction between superconductor, iron, and rotor}

We consider full magnetic interaction of the superconductor, the permanent magnets in the rotor, and the iron as follows. 

First, we assume uniform magnetization of the permanent magnets in the rotor. We also assume uniform $\vJ$ in all the satator coils at the beginning. Then, we solve $\vM$ in the iron considering the applied flux density, $\vB_a$, from the permanent magnets and the stator coils. Afterwards, we solve the current density in the superconducting coils (solid rectangles in figure \ref{f.motor}(top)) considering the applied vector potential, $A_a$, from the permanent magnets and the iron. Finally, we iterate until the difference in all components of both $\vM$ and $J$ is below a certain tolerance.

\begin{figure}[tbp]
{\includegraphics[trim=0 0 0 0,clip,width=\figwidth]{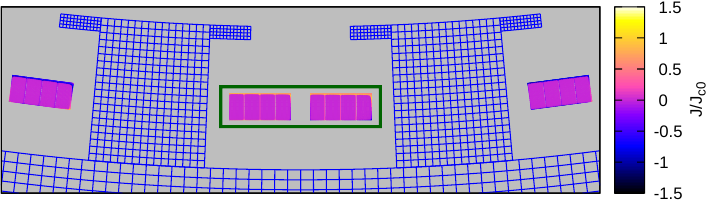}}%
\captionof{figure}{Current density at the peak of the current of coils 1 and 2 (half-coils 1 to 4) for the coupled-at-ends case. Above, $J_{c0}=J_c(B=0,\theta=0,T=40\ {\rm K})$. The green frame denotes the zoom in figure \ref{f.Jzoom}. In the maps, the superconductor thickness has been expanded for better visialization.}%
\label{f.J}%
\end{figure}

Figure \ref{f.M} shows an example of the solution of $\vM$ and figure \ref{f.J} shows a solution of $J$.

\section{Benchmark with $T-A$ formulation in COMSOL}

\begin{figure}[tbp]
{\includegraphics[trim=0 0 0 0,clip,width=\figwidth]{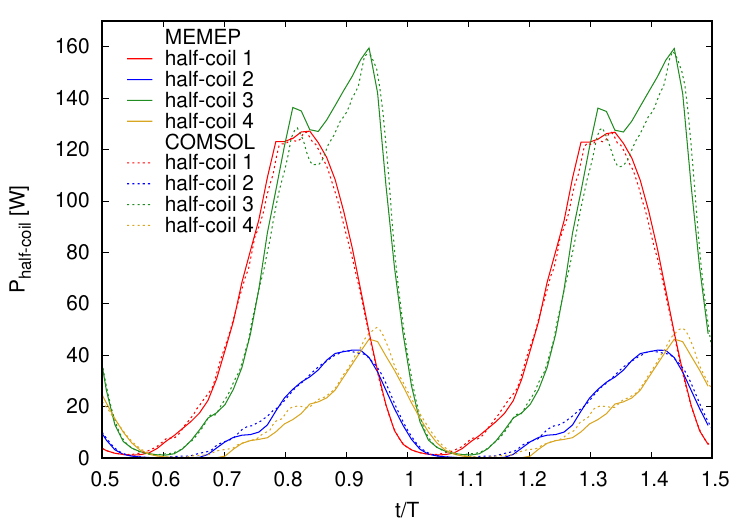}}%
\caption{The instantaneous power loss in each half-coil for MEMEP and COMSOL agree to each other, supporting the validity of MEMEP. \label{f.bench}}%
\end{figure}

In order to confirm the correctness of the MEMEP numerical method, we benchmarked the motor results with the $T-A$ formulation of the Finite Element Method (FEM) using COMSOL software \cite{benkelT2020IES, huberF2022SST}. For this case, we considered 2 superconducting coils only, in addition to the coils with uniform current density. We also assumed the ``uncoupled'' configuration. The results show good agreement of the instantaneous AC loss at each half-coil (figure \ref{f.bench}). The remaining small discrepancy is because of different meshing of both superconductor and iron yoke.

\section{Results}

\subsection{Effect of coupling}

\subsubsection{Simplified coupling}

\begin{figure}[tbp]
{\includegraphics[trim=0 0 0 0,clip,width=\figwidth]{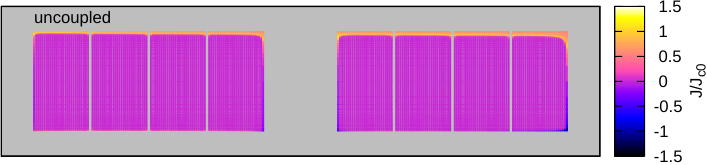}}\\%
{\includegraphics[trim=0 0 0 0,clip,width=\figwidth]{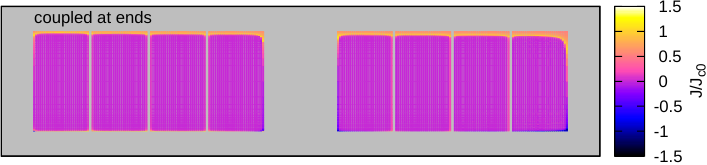}}\\%
{\includegraphics[trim=0 0 0 0,clip,width=\figwidth]{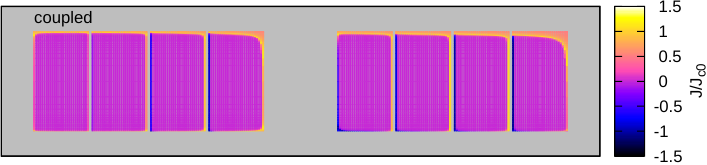}}%
\caption{Current density at half-coils 2 and 3 for the three coupling configurations of figure \ref{f.coupl}.
\label{f.Jzoom}}%
\end{figure}

\begin{figure}[tbp]
{\includegraphics[trim=0 0 0 0,clip,width=\figwidth]{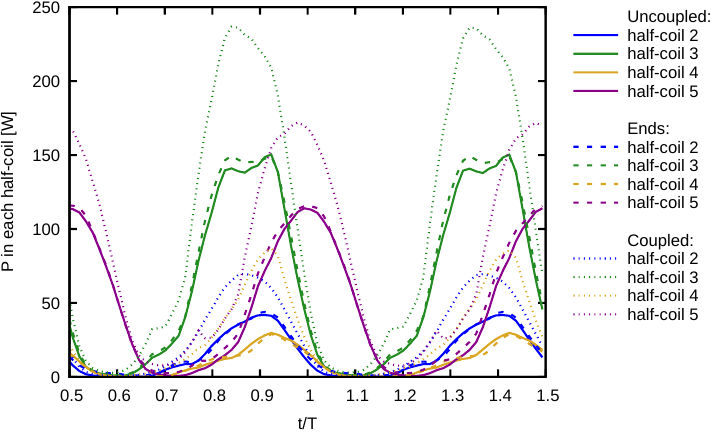}}%
\caption{The instantaneous power loss at each half-coil in the ``coupled'' case of figure \ref{f.coupl} is much higher than ``coupled at ends'' (or ``ends'') and ``uncoupled''. Shown is the AC loss for half-coils 2 to 5 in figure \ref{f.motor}. The AC loss in the other half-coils is the same but with a certain phase shift. \label{f.P}}%
\end{figure}

\begin{figure}[tbp]
{\includegraphics[trim=0 0 0 0,clip,width=\figwidth]{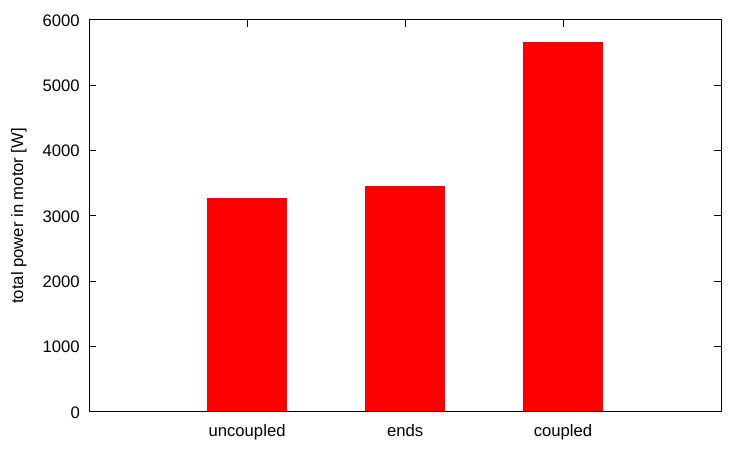}}%
\caption{The AC loss for the realistic coupling scenario ``coupled-at-ends'' (or ``ends'') of figure \ref{f.coupl} is below 0.18 \% of the motor rated power. \label{f.Ptot}}%
\end{figure}

Next, we analyze the three coupling limits of figure \ref{f.coupl}. Here, we focus on the current density at the peak of the current. The current density maps for the uncoupled case show that the current density concentrates almost entirely on the closest side of the rotor (top side in figures \ref{f.J} and \ref{f.Jzoom}). The current density for the coupled-at-ends case is almost the same as uncoupled. The reason is that the magnetic flux between parallel tapes in the conductor due to one coil self-field practically vanishes. Indeed, it vanishes completely if the bore is infinite. In addition, part of the magnetic flux from the rotor is also canceled because each half of the same coil is submitted to the magnetic flux density of a magnet of opposite polarity. Then, any residual flux imbalance is due to either neighboring coils, the iron core, or the rotor. This explains the slightly higher current density at the rightmost and leftmost tapes of each turn. For the uncoupled configuration, there is substantially higher current penetration from the top. In addition, the rightmost and leftmost tapes of certain turns are saturated with $|J| \approx J_c$.

We compute the instantaneous AC loss, $P$, from the power density $p=J\cdot E(J)$ at each cell, and then summing the contribution from all cells. Thanks to the motor symmetry, it is enough to compute the time-average power loss in two neighboring slots and multiply by the number of slots. However, in order to accurately take coupling effects into account we need to compute full coils, which have each half-coil in different slots. We have made computations from 3 to 5 coils and we have seen that the difference in the results is negligible. Therefore, it is enough to consider 3 full coils to accurately calculate the AC loss in two consecutive slots for any coupling scenario. For 3 coils, only half-coils 2,3,4,5 in figure \ref{f.motor} are considered, although $J$ is solved in all 6 coils. 

The instantaneous power at each coil in figure \ref{f.P} shows that the realistic ``coupled-at-ends'' configuration presents almost the same AC loss of the ideal ``uncoupled'' case, with slightly higher loss for ``coupled-at-ends''. The reason is that almost all the increase in screening currents for ``copuled-at-ends'' occurs at the rightmost and leftmost tapes of each turn with $|J|$ below the local $J_c$, and hence $E$ practically vanishes, and so does $p$. However, the ``coupled'' case presents much higher AC loss than ``coupled'' or ``coupled-at-ends'', since $|J|$ overcomes $J_c$ in almost the whole section of the rightmost and leftmost tapes of several turns.

The time average power loss in the whole stator for ``coupled'' is almost twice that of ``uncoupled'', while ``coupled-at-ends'' is only slightly higher than ``uncoupled'' (figure \ref{f.Ptot}). For ``coupled-at-ends'', this AC loss is around 3500 W, which is less than 0.18 \% of the rated power. Then, the motor is highly electrically efficient. Although this heat generation is considerable from the cryogenic point of view, it is feasible thanks to the large liquid hydrogen reservoir at 20 K, which acts as ultimate heat sink.

The relatively low AC loss is due to the stacking effect. When the applied magnetic flux density is well below the saturation flux density of a stack of tapes, the AC loss is well below the AC loss of a single tape \cite{pardoE2003PRB}. Similarly, the AC loss in a coil submitted to a transport current, decreases with the coil critical current. Since the latter increases with the number of tapes, the AC loss decreases with the number of tapes.

\subsubsection{Coupling by finite resistances at terminals}

\begin{figure}[tbp]
{\includegraphics[trim=0 0 0 0,clip,width=\figwidth]{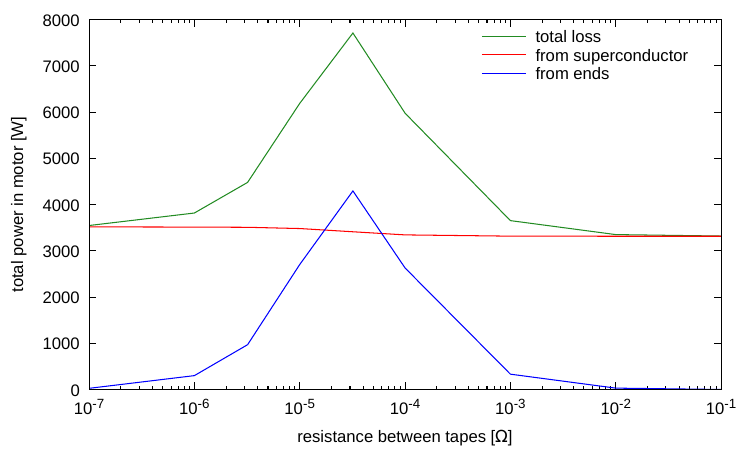}}%
\caption{AC loss of the intermediate configuration between ``coupled-at-ends'' and ``uncoupled'' of figure \ref{f.couplR}(left). For the studied motor, most of the coupling current flows from the two extreme tapes of the cable, and hence the AC loss will be practically the same as realistic terminal connection of figure \ref{f.couplR}(right). \label{f.PR}}%
\end{figure}

Next, we study the smooth transition between ``uncoupled'' and ``coupled-at-ends'' by placing finite resistances close to the terminals (figure \ref{f.couplR}(a)). For a range of resistance between $10^{-6}$ and $10^{-3}$ $\Omega$, there is a significant AC loss contribution from the resistances (figure \ref{f.PR}), which could double the total AC loss. However, for end resistances beyond this range, the end resistance contribution is negligible.

In this configuration, almost all net current flowing between tapes is between the leftmost and rightmost tapes. Then, the AC loss scenario that we have considered is almost the same as soldering the tapes directly to a current terminal (figure \ref{f.couplR}(b)), but with end resistances following (\ref{Rl}). To this AC loss contribution, we should also add the DC loss contribution, $P_{DC}=I_{\rm rms}^2(R_l/n)$. Then, we should aim to the lowest possible $R_l$. If $R_s<1$ $\mu\Omega$ ($R_l<13.5$ $\mu\Omega$), then $P_{DC}<12$~W; and hence both AC and DC contributions to the power loss at the terminals are negligible.

\subsection{Effect of misalignment}

\begin{figure}[tbp]
{\includegraphics[trim=0 0 0 0,clip,width=\figwidth]{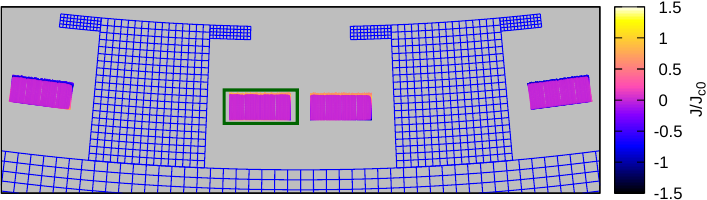}}\\%
{\includegraphics[trim=0 0 0 0,clip,width=\figwidth]{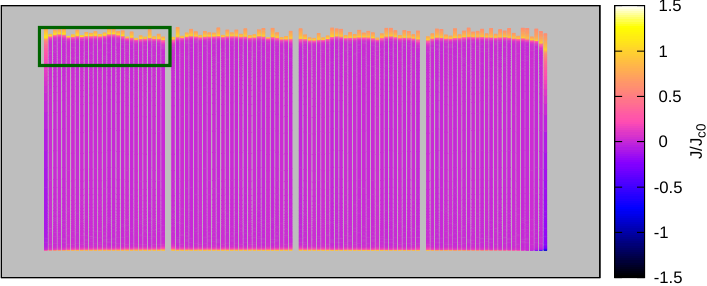}}\\%
{\includegraphics[trim=0 0 0 0,clip,width=\figwidth]{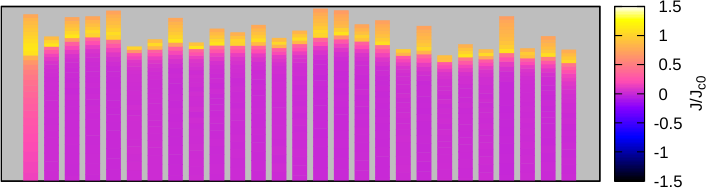}}%
\caption{Current density in the superconducting coils when considering random tape misalignment within the $\pm$100 $\mu$m range. The green frame delimits the zoomed area for the next map. \label{f.Jrand}}%
\end{figure}

\begin{figure}[tbp]
{\includegraphics[trim=0 0 0 0,clip,width=\figwidth]{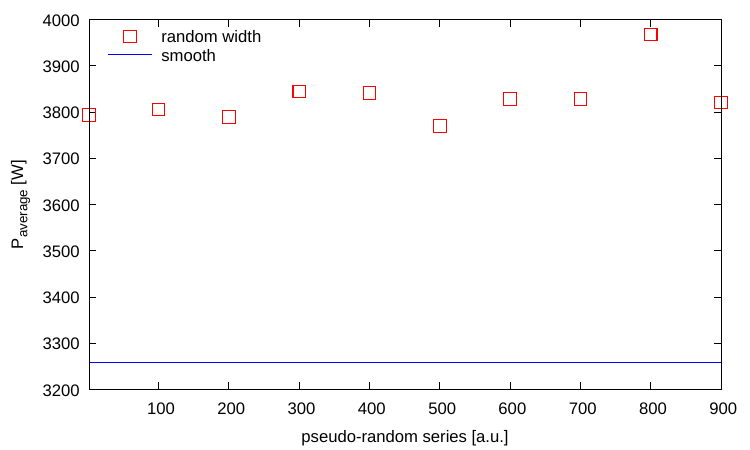}}%
\caption{ Time-average AC loss for 10 different series of random width distribution for the ``uncoupled'' configuration. \label{f.Prand}}%
\end{figure}

The low AC loss of the rotor relies on the stacking effects. Then, misalignment between tapes could increase AC loss, compromising the stacking effect. In this section, we study the impact of random misalignment between tapes. Here, we present the results with misalignment from the side closest to the rotor, since we have checked that misalignment from the other side does not appreciably increase the AC loss. 

We consider a quasi-random distribution with no correlation between different coils (figure \ref{f.Jrand}). The variation of the random tape is $\pm 0.1$ mm, since this is comparable to both experimental misalignment in coil winding and tape width variation. For random number generation, we used the \verb|rand()| function in C++, which provides random numbers with almost equal probability.

\subsection{Degradation at edges}

\begin{figure}[tbp]
{\includegraphics[trim=0 0 0 0,clip,width=\figwidth]{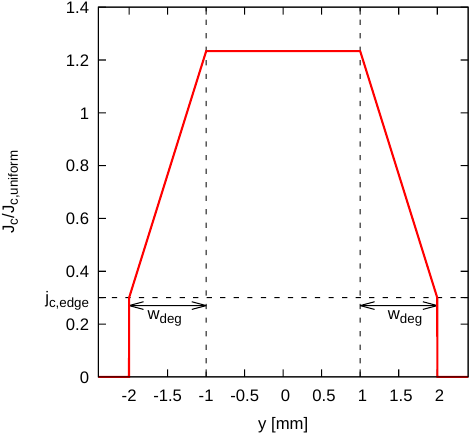}}%
\caption{Assumed generic non-uniformity of the critical current density. The $J_{c,{\rm uniform}}$ value corresponds to the experimental data of figure \ref{f.JcBth}. \label{f.Jcnor}}%
\end{figure}

\begin{figure}[tbp]
{\includegraphics[trim=0 0 0 0,clip,width=\figwidth]{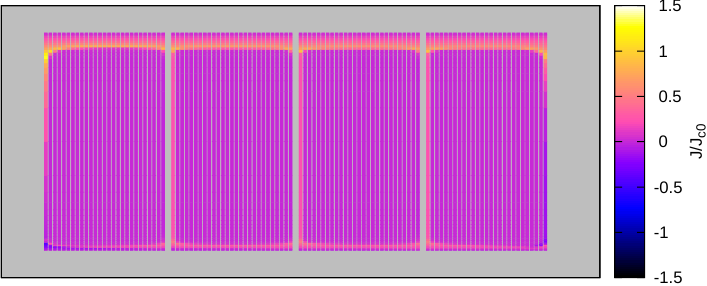}}%
\caption{Current density at half-coil 2 with non-uniform $J_c$ for $j_{c,{\rm edge}}=0$ and $w_{\rm deg}=0.4$ mm (parameter definition in figure \ref{f.Jcnor}).\label{f.Jdeg}}%
\end{figure}

\begin{figure}[tbp]
{\includegraphics[trim=0 0 0 0,clip,width=\figwidth]{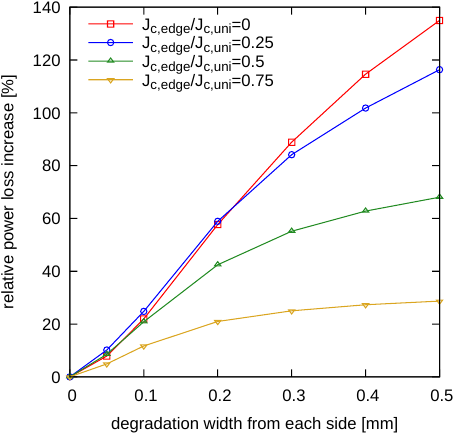}}%
\caption{The AC loss increases with the depth of $J_c$ degradation but decreases with $J_c$ at the edge (generic $J_c$ non-uniformity in figure \ref{f.Jcnor}). \label{f.Pavdeg}}%
\end{figure}

Loss reduction due to stacking effect could also be compromised by degradation close to the tape edges. Here, we consider degradation of $J_c$ following the distribution of figure \ref{f.Jcnor}, which assumes symmetric degradation from both sides, linear decay of $J_c$ from the center to the edge, and the same critical current as assuming uniform $J_c$ for a given magnetic field, $B$, its angle, $\theta$, and temperature, $T$. Thus, $J_c/J_{c,{\rm uniform}}$ in figure \ref{f.Jcnor} represents a factor that is multiplied to $J_c(B,\theta,T)$ in order to obtain the local $J_c$.

The results show a larger penetration of $|J|\approx J_c$ from the top (figure \ref{f.Jdeg}, which is consistent with decreased $J_c$. Since most of $J$ experiences reduced $J_c$ compared to the uniform $J_c$ case, the AC loss increases (see figure \ref{f.Pavdeg}). The increase in AC loss is above factor 2 for extreme cases of degradation ($J_{c,{\rm edge}}=0$ and large degradation depth). However, degradation is tolerable if it is restricted to less than 100 $\mu$m from the edge, where the loss increase is below 25 \%. Therefore, degradation is not expected to highly impact the AC loss, unless it propagates relatively deeply within the tape. A conclusion is that in order to reduce AC loss, it is not sufficient that REBCO tapes present high critical current, but also it should not be degraded at the edges. Therefore, careful tape characterization in this respect is needed.

\section{Conclusion}

This article has presented a novel modeling method (MEMEP) for the AC loss in superconducting rotating machines, which could also be applied to normal conducting machines. We applied this method for a case study of a propulsion motor for a hydrogen-electric aircraft, which contains a superconducting stator.

We have found that the dissipated power from MEMEP agrees with finite-element method (FEM) computations with COMSOL, which supports the validity of MEMEP. Then, we have analyzed the AC loss under several electric coupling scenarios. For the realistic configuration of coupled-at-ends, we have found that the AC loss is below 0.18 \% of the motor rated power. However, the heat generated is still considerable from the cryogenic point of view (around 3500 W), and hence heat extraction at the relatively low operation temperature (40 K) should be made with care. 

The main reason of low loss in the REBCO winding is the stacking effect. In this article, we have checked that typical tape misalignment does not compromise loss reduction by the stacking effect. Tape degradation at the edges could substantially increase the AC loss, although the increase is moderated if degradation does not reach depths higher than around 100 $\mu$m. Then, tape homogeneity should be minimized for motor applications. Soldering all tapes to each other across their length will incur to a high AC loss increase, as inferred from the ``coupled'' scenario.

Future work will be dedicated to show that the metallic layers in the superconductor do not significantly increase the AC loss. Work will also be done regarding the electrothermal behavior.

\section{Acknowledgments}

Funded by the EU NextGenerationEU through the Recovery and Resilience Plan for Slovakia under the project No. 09I04-03-V02-00039, the Slovak Research and Development Agency (project No. APVV-24-0654), and ``Vedecká grantová agentúra MŠVVaM SR a SAV'' (VEGA) (project No. VEGA 2/0098/24).

\section*{Declaration}

No artificial intelligence has been used in writing this article or preparing any graph.

\appendix

\section{The Minimum Electro-Magnetic Entropy Production (MEMEP) method}
\label{s.MEMEP}

This section details the key aspecs of the MEMEP method for electromagnetic modelling, outlined in section \ref{s.scmod}. 

In general, the current density, $\vJ$, in any conductor with any consititutive $\vE(\vJ)$ relation follows equation (\ref{EAphi}). Discretizing in time, (\ref{EAphi}) becomes
\begin{equation} \label{EDAphi}
\vE(\vJ)=-\frac{\vA[\Delta\vJ]}{\Delta t}-\frac{\Delta\vA_a}{\Delta t}-\nabla\phi,
\end{equation}
where $\Delta t$ is a finite time increment, $\Delta\vJ\equiv \vJ(t)-\vJ(t-\Delta t)$, and $\Delta\vA_a=\vA_a(t)-\vA_a(t-\Delta t)$. In \cite{pardoE2017JCP} we have found that (\ref{EDAphi}) is the Euler equation of the following functional, and hence we can find $\vJ$ by minimizing the functional at each time step, starting from a known $\vJ$ at $t=0$ [it could be $\vJ(t=0)=0$]. The functional is
\begin{eqnarray} \label{L3D}
L[\vJ] & = & \int_\Omega\dvol \biggl\{ \frac{1}{2}\vDJ\cdot\frac{\vDA[\vDJ]}{\Delta t} + \vDJ\cdot\vDA_a \nonumber \\
&& + U(\vJ) +\nabla\phi\cdot\vJ \biggr\},
\end{eqnarray}
where $U(\vJ)$ is defined as
\begin{equation}
U(\vJ)\equiv \int_0^{\vJ}\dif\vJ'\cdot \vE(\vJ').
\end{equation}

For problems with length $l$ in the $z$ direction much larger that the width, we can neglect the contribution the ends to $L$. Then, $\vJ\approx J\ve_z$ and $\vE\approx E\ve_z$, where $\ve_z$ is the unit vector in the $z$ direction. With no resistances at the terminals, the functional above reduces to \cite{pardoE2015SST}
\begin{equation} \label{L2Dgen}
L[\vJ]=L_l[\vJ]+L_e[\vJ]
\end{equation}
with
\begin{eqnarray}
L_l[\vJ] & = & l\int_{\Omega_s}\dsec \biggl\{ \frac{1}{2}\Delta J\frac{A[\Delta J]}{\Delta t} + \Delta J\frac{\Delta A_a}{\Delta t} \nonumber \\
&& + U(J) \biggr\} \label{L2D} \\
L_e[\vJ] & = & - \sum_{\alpha=1}^{n_{\rm co}} \sum_{i=1}^nV_{\alpha i}I_{\alpha i},
\end{eqnarray}
where $\Omega_s$ is the whole cross-section where there is $J$, $n_{\rm co}$ is the number of coils, $n$ is the number of tapes in parallel in the cable of each coil, $V_{\alpha i}$ is the total voltage between terminals of tape $i$ and $I_{\alpha i}$ is the net current in cable $i$. For the coupled and coupled-at-ends configuration, $V_{\alpha i}=V_{\alpha}$ for all tapes, and hence the last term reduces to $-\sum_{\alpha=1}^{n_{\rm co}} V_\alpha I_\alpha$, where $I_\alpha$ is the total transport current of coil $\alpha$. For ``uncoupled'', $I_{\alpha i}=I_\alpha/n$, which simplifies that term into $-\sum_{\alpha=1}^{n_{\rm co}}I_\alpha \sum_{i=1}^n V_{\alpha i}/n$. For all these three cases, the last term does not depend on the particular $J(x,y)$ distribution, as long as the net current at each coil, $I_\alpha$, is maintained. Thus, the term $L_e$ in (\ref{L2Dgen}) can be dropped from the minimization.

For partial coupling, we also need to take the contribution to $U(\vJ)$ in (\ref{L3D}) into account for the resistances. Neglecting screening currents at the end resistances, for one coil this term is $(1/2)\sum_{i=1}^{n-1}I_{{\rm l}i}^2R_{{\rm l}i}$ for each terminal, where $I_{{\rm l}i}$ is the loop current of mesh $i$ and $R_{{\rm l}i}$ is the end resistance at mesh $i$. For one coil, the functional still follows (\ref{L2Dgen}) with $L_l$ from (\ref{L2D}) but with 
\begin{equation} \label{LeRin}
L_e[\vJ]=\sum_{i=1}^{n-1}I_{{\rm l}i}^2R_{{\rm l}i}
-2\sum_{i=1}^{n-1}I_{{\rm l}i}V_{{\rm R}i}
-\sum_{i=1}^nV_iI_i,
\end{equation}
where $R_{{\rm R}i}$ is the voltage drop at one resistance at loop $i$. From current conservation at the junctions at figure \ref{f.couplR}(a), the current at each tape follows
\begin{equation}
I_i=
\begin{cases} 
			I/n+I_{{\rm l}i}-I_{{\rm l}i-1} & \text{if $1<i<n$}\\
			I/n+I_{{\rm l}i} & \text{if $i=1$} \\
			I/n-I_{{\rm l}i-1} & \text{if $i=n$}
\end{cases}.
\end{equation}
Using this, and that by Kirshoff's loop law $V_{i+1}=V_i+2V_{{\rm R}i}$, we find that (\ref{LeRin}) becomes
\begin{equation}
L_e[\vJ]=\sum_{i=1}^{n-1}I_{{\rm l}i}^2R_{{\rm l}i}
-\frac{I}{n}\sum_{i=1}^nV_i.
\end{equation}
For current constraint (given $I$), we can drop the last term because it is independent on the particular $\vJ(\vr)$ distribution.

For the case with resistances at the terminals of figure \ref{f.couplR}(b), we also need to add a term from the resistance. For that case, the functional for one coil is again (\ref{L2Dgen}) with $L_l$ from (\ref{L2D}) but with $L_e$ given by 
\begin{equation}
L_e=\sum_{i=1}^{n}I_i^2R_i-\sum_{i=1}^nI_i(V_i+2V_{{\rm R}i})=\sum_{i=1}I_i^2R_i-VI,
\end{equation}
where the sums are done among all parallel tapes in the cable and $V_{{\rm R}i}$ is the voltage in one resistance at tape $i$ at the terminals. Above, we have used that $V=V_i+2V_{{\rm R}i}$ for all $i$, since the superconducting tape is in series with the resistances at the terminals. For current constraints ($I$ is given), we can again omit the $VI$ term in the equation above.

\section{Non-linear mesh}

In this article, we use non-linear mesh across the width of the tape, while we only consider one element in the superconductor thickness. This non-linear mesh generates small cells close to the tape edges, where most of the AC loss concentrates, and relatively large cells at the center, where the AC loss practically vanishes. The mesh is divided using the following meshing function
\begin{equation}
f(y')=\sinh(k'y')/C,
\end{equation}
where 
\begin{equation}
C\equiv \sinh k',
\end{equation}
$y'$ is the coordinate along the tape width in a Cartesian coodinate system with origin at the tape center divided by the tape half-width ($w/2$), and $k'$ is a paramter determining the non-uniformity of the mesh. In this way, we make the non-linear mesh by dividing the range of $f$, which is from $-1$ to 1, into $n_y$ values and obtaining the node positions following that
\begin{eqnarray}
f_i & = & i\Delta f-1 \\
y'_i & = & \frac{1}{k'}{\rm asinh}(f_iC),
\end{eqnarray}
where $\Delta f=2/n_y$ and $i$ is the node index (it ranges from 0 to $n_y$ included). Instead of using $k'$ as parameter it is more practical to use the ratio of the cell size at the center of the rectangle with that of the edge, which approximately corresponds to the slope of $f$ at the edge divided by the slope of $f$ at the center. This ratio, $r_f$, is
\begin{equation}
r_f=\cosh k'.
\end{equation}
Typical computations use $r_i=32$ and $n_y=38$. This has a cell size at the edge similar to $n_y=160$ for uniform mesh. Accurate calculations, such as those for tape misalignment, use $r_f=32$ and $n_y$=77 (cell size at edge equivalent to 320 elements with uniform mesh).


\newpage

\newpage

\end{document}